\documentclass[prl,twocolumn]{revtex4}

\usepackage{graphicx}

\begin{document}

\title{Control of the persistent currents
in two interacting quantum rings through the Coulomb interaction and
inter-ring tunneling}
\author{L. K. Castelano$^{1}$}
\author{G.-Q. Hai$^{1}$}
\email{hai@ifsc.usp.br}
\author{B. Partoens$^{2}$}
\author{F. M. Peeters$^{2}$}
\affiliation{$^{1}$Instituto de F\'isica de S\~ao Carlos,
Universidade de S\~ao Paulo, 13560-970, S\~ao Carlos, SP, Brazil
\\
$^{2}$Department of Physics, University of Antwerp,
Groenenborgerlaan 171, B-2020 Antwerp, Belgium}

\begin{abstract}
The persistent current in two vertically coupled quantum rings
containing few electrons is studied. We find that the Coulomb
interaction between the rings in the absence of tunneling affects
the persistent current in each ring and the ground state
configurations. Quantum tunneling between the rings alters
significantly the ground state and the persistent current in the
system.
\end{abstract}

\maketitle


\section{Introduction}
Mesoscopic metallic rings and nanoscopic quantum rings (QRs) are
known for the Aharonov-Bohm effect and its persistent
current\cite{AB} where the phase rigidity of the electron wave leads
to quantum oscillations in the current. Recently, oscillatory
persistent current was detected in self-assembled InAs/GaAs
semiconductor QRs having a single electron\cite{Kleemans}.
Experimentally there have been a few approaches to fabricate coupled
rings and little effort has been done on the system of vertically
coupled QRs\cite{granados}. Due to the ring geometry, these coupled
ring complexes open a new route for measurement of quantum
interference effects and for novel many-body
states\cite{lkc,malet,szafran,he}. Two vertically coupled quantum
rings (CQRs) form a new type of artificial molecule. The inter-ring
distance and the tunneling strength together with the ring radius
are new tunable parameters providing more degrees of freedom to
modulate and to control the electronic structure and the persistent
current of these ring shaped artificial molecules. In the present
work, within the current spin-density functional theory
(CSDFT)\cite{vignale,vief}, we study the persistent currents in the
CQRs of few-electrons. The two coupled rings exhibit new molecular
many-body states. The tunability of the inter-ring distance and the
inter-ring tunneling strength leads to a very rich variety of
many-electron ground states and persistent current spectra. The
persistent current and magnetization in a single ring have been
studied extensively where the external magnetic field is the only
parameter to alter the current at fixed ring radius. Here, we show
that the persistent current in the CQRs depends on the inter-ring
distance and the tunneling strength between the two rings. Tunneling
between the rings permits the exchange interaction between them
inducing new quantum states that alter significantly the persistent
current. This opens new ways to control the persistent current in
quantum rings.

\section{Theoretical model}
We firstly study the inter-ring Coulomb interaction effects on the
persistent current in two interacting QRs without tunneling. We
model the QRs by a displaced parabolic potential
$V(r)=\frac{1}{2}m^{\ast }\omega _{0}^{2}(r-r_{0})^{2}$ in the
$xy$-plane, where $\mathbf{r}=(x,y)=(r,\theta ) $, $\omega _{0}$ is
the confinement frequency and $r_{0}$ is the radius of the ring. The
two stacked identical rings are separated by a distance $d$ in the
$z$ direction. They interact with each other through the Coulomb
potential. A homogeneous magnetic field $\mathbf{B}=B\mathbf{e}_z$
is applied perpendicularly to the $xy$-plane, which is described by
the vector potential $\mathbf{A}=Br\mathbf{e}_{\theta}/2$ taken in
the symmetric gauge. The Kohn-Sham orbitals $\psi _{jnm\sigma
}(\mathbf{r})=\exp (-im\theta )\phi _{nm\sigma }(r)Z_j(z)$ are used
to express the density and ground state energy, where $\sigma
=\uparrow $ or $\downarrow $ is the $z$ component of the electron
spin and $\omega_c=eB/m^\ast c$ is the cyclotron frequency. The
total density in the rings is $\rho (\bf r)=\sum_{\sigma
}\sum_{n,m}^{N_{\sigma }}|\phi _{nm\sigma }(\bf r)|^{2}$. In the
$z$-direction this density is approximated by a $\delta$-function in
each ring. The inter-ring Coulomb potential is given by
\begin{equation}
V_{H}^{\text{inter}}(r)=\int d\mathbf{r^{\prime }}\frac{e^{2}\rho (r^{\prime
})/2}{\varepsilon |\mathbf{r}-\mathbf{r^{\prime }}+\mathbf{d}|},
\end{equation}%
with the inter-ring distance $d=|\mathbf{d}|$. The
exchange-correlation potential is considered within the
local-density approximation.

In the CSDFT all the quantities are functionals depending on the
spin-up ($\rho^\uparrow$) and spin-down ($\rho^\downarrow$)
densities, and the vorticity $\mathbf{\mathcal{V}}(r)=\frac{m^\ast
c}{e}\nabla\times \mathbf{j}_p(r)/\rho(r)$. Therefore the
exchange-correlation scalar and vector potential can also be written
as a functional of these quantities\cite{vignale,bart}. The
paramagnetic current density $\mathbf{j}_p (r)$ can be explicitly
written as
\begin{equation}\label{parcur}
 \mathbf{j}_p (r)=-\frac{\hbar}{m^\ast r}\mathbf{e}_\theta\sum_{\sigma
}\sum_{n,m}^{N_{\sigma }}m|\phi _{nm\sigma }(r)|^{2}.
\end{equation}
The persistent current density, which is a measurable quantity, is
given by $\mathbf{j}(r)=\mathbf{j}_p(r)+(e/m^\ast
c)\rho(r)\mathbf{A}(r)$, where the second part corresponds to the
diamagnetic current density. The persistent current is obtained by
$I=\int j(r) dr =I_p+I_d$. We confirm that the magnetization in both
the single and coupled QRs is proportional to the persistent
current.

When inter-ring quantum tunneling is permitted, the electronic
states in different rings are coupled and the CQRs form an
artificial molecule. There are bonding and antibonding states with
an energy splitting $\Delta_{\rm SAS}$ between the two levels.
Tunneling occurs through a potential barrier between the rings in
the z-direction $V(z)$ which is modeled by two coupled symmetric
GaAs quantum wells of width $120$ \AA \, and a barrier of $250$ meV
height between them. Within this model, the energy splitting is
given by $\Delta_{\rm SAS} =22.86$ $\exp [-22.42 d(r_0)]$ meV as a
function of the inter-ring separation\cite{lkc}. For the CQRs with
inter-ring tunneling, each GS can be labeled by three quantum
numbers $(S_{z},M_{z},I_{z})$: the total spin $S_{z}$, total angular
momentum $M_{z}$ and the isospin quantum number $I_{z}$.

\begin{figure}
\includegraphics[angle=-90,scale=0.35]{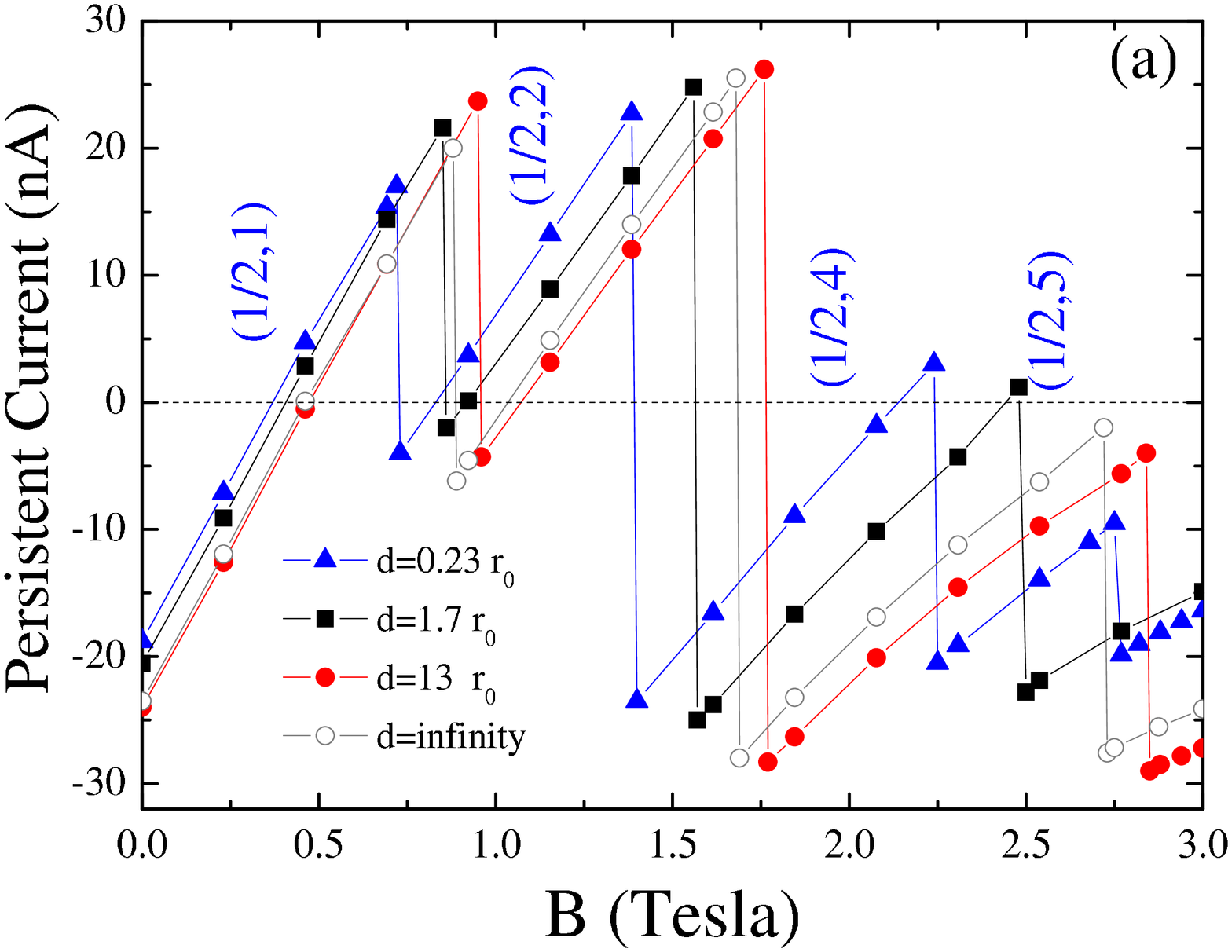}
\includegraphics[angle=-90,scale=0.35]{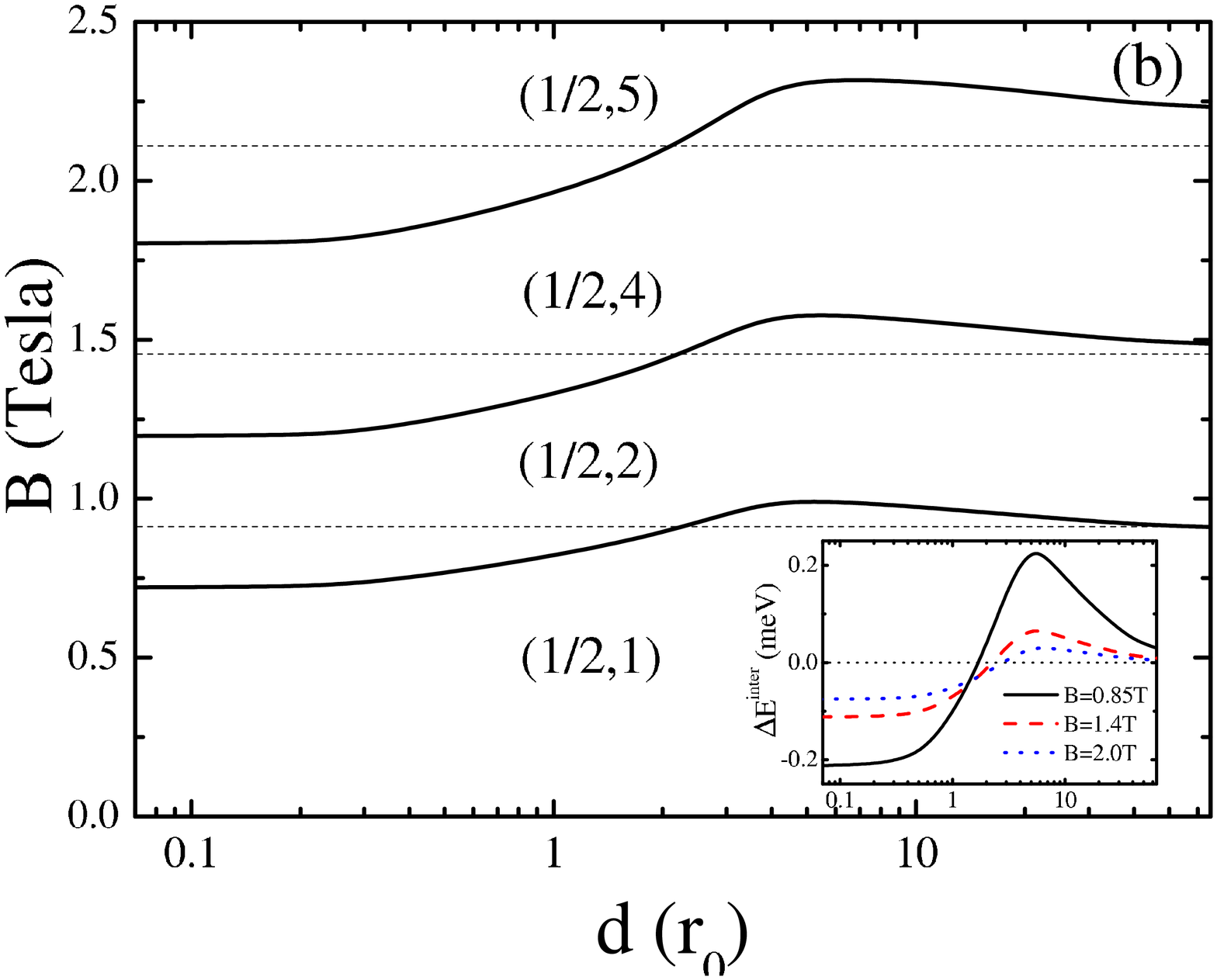}
\includegraphics[angle=-90,scale=0.35]{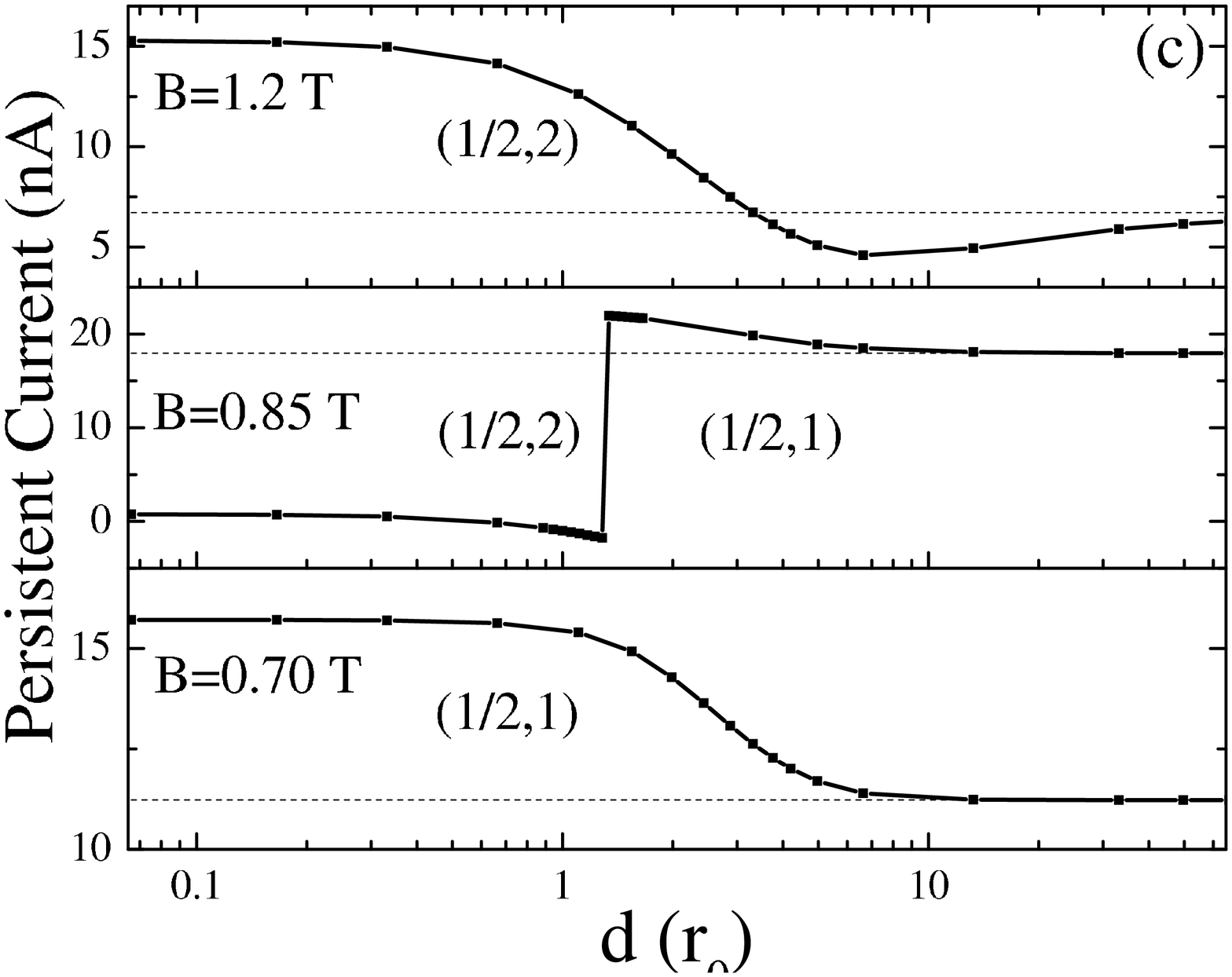}
\caption{(Color online) (a) The persistent current as a function of
magnetic field at different $d$ in two interacting QRs containing
each 3 electrons without inter-ring tunneling. (b) The phase diagram
in the B-d plane. The dashed horizontal lines indicate the
transitions in a single ring. Each phase is indicated by two quantum
numbers ($S_z,M_z$). The inset shows the energy difference $\Delta
E^{\rm inter}$ of the inter-ring Coulomb energy between the
different configurations at a fixed magnetic field. (c) Variation of
the persistent current in the rings as a function of $d$ at B=0.7,
0.85, and 1.2 T. The horizontal lines are the current for the single
ring case. }\label{fig.1}
\end{figure}

\section{Numerical Results}
In the calculations, we consider GaAs quantum rings with a ring
radius $r_0=2$ a$_0$ and a confinement energy $\hbar\omega_0=5$ meV,
where $a_0=\sqrt{\hbar/m^*\omega_0}$. The inter-ring Coulomb
interaction effects are studied by varying the distance $d$ between
the two rings. The ground state (GS) of a single ring can be
characterized by two quantum numbers: the total spin $S_z$ and the
total angular momentum $M_z$. The persistent current is obtained for
two interacting rings without tunneling between them,
containing each 3 electrons, at different
inter-ring distances. As shown in Fig.~1(a), the current is an
oscillation function of the magnetic field with semi-linear
segments. The different segments appear because $M_z$ increases with
increasing magnetic field leading to a GS configuration (or phase)
transition as indicated in the figure. At small magnetic fields, the
GS phase of a 3 electron QR is ($S_z,M_z$)=($1/2,1$). With
increasing magnetic field, the total angular momentum increases as
$M_z$=1,2,4, and 5, but the total spin remains $S_z$=1/2. We observe
that the position of the jumps in the current is not a monotonous
function of $d$. Fig. 1(b) presents a phase diagram of the ground
states of the QRs in the $B$ vs. $d$ plane. The horizontal dotted
lines indicate the GS transitions of a single ring. The GS depends
on the inter-ring distance $d$ and this dependence is stronger when
$d$ is close to the diameter of the rings (2r$_0$). It results
mainly from the inter-ring Coulomb energy which is different for the
different GS configurations (different $M_z$) as shown in the inset
of Fig.~1(b). The energy difference $\Delta E^{\rm inter}$ of the
inter-ring coulomb energy $E^{\rm inter}_{\rm S_z,M_z}$ between the
different configurations ($S_z,M_z$) and ($S'_z,M'_z$) is plotted as
a function of $d$: $\Delta E^{\rm inter}=E^{\rm inter}_{\rm
S_z,M_z}-E^{\rm inter}_{\rm S'_z,M'_z}$. The three curves are
$E^{\rm inter}_{1/2,2}-E^{\rm inter}_{\rm 1/2,1}$ at $B=0.85$ T,
$E^{\rm inter}_{1/2,4}-E^{\rm inter}_{\rm 1/2,2}$ at $B=1.4$ T,
$E^{\rm inter}_{1/2,5}-E^{\rm inter}_{\rm 1/2,4}$ at $B=2.0$ T. The
energy difference is zero for $d\approx 2$ r$_0$. In comparison with
the single ring, the phase transition occurs at lower (larger)
magnetic fields when $d<2$r$_0$ ($d>2$r$_0$). As a consequence, we
found the remarkable behavior that the inter-ring Coulomb
interaction affects significantly the persistent current. Fig.~1(c)
shows that, at $B=$0.7 and 1.2 T where the GS configurations are the
same for all $d$, the persistent current decreases with increasing
$d$ till a minimum and then increases slowly to its limit value of
an isolated ring. However, variation of the inter-ring distance at
fixed magnetic field can also induce the GS phase transition and,
consequently, a jump in the persistent current. This is the case of
$B=0.85$ T in Fig.~1(c) where the GS configuration changes from
(1/2,2) to (1/2,1) at $d=$1.2r$_0$ leading to an abrupt jump of the
persistent current from -4 to 22 nA.
\begin{figure}[t]
\includegraphics[angle=-90,scale=0.35]{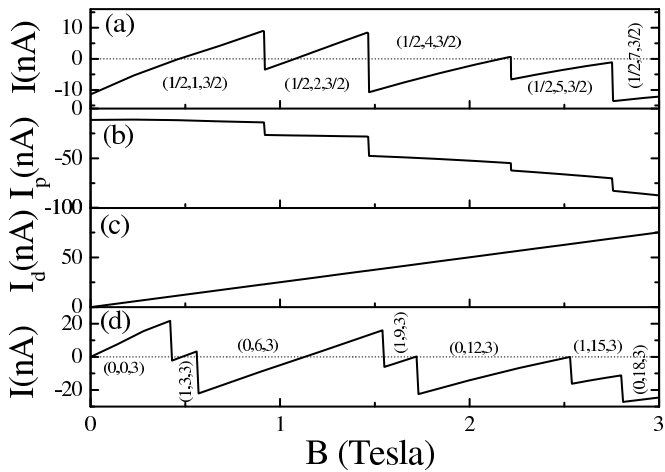}
\caption {(a) The total persistent current, (b) the paramagnetic and
(c) the diamagnetic current as a function of magnetic field in the
CQRs of $d=0.1$r$_0$ and with $N=$3 electrons. (d) The same as (a)
but now for $N=$6. The different phases are indicated by
$(S_{z},M_{z},I_{z})$.}\label{fig.2}
\end{figure}

The CQRs form an artificial molecule when inter-ring quantum tunneling
is permitted. Fig.~2(a) shows the persistent current as a function
of magnetic field in such an artificial molecule with a total of 3
electrons in the strong tunneling regime for $d=0.1$~r$_0$.
In this regime, the CQRs behave as a single one
because only the bonding states are occupied. This is evident by the
isospin quantum number $I_z$ which is always 3/2. The spin and the
angular momentum ($S_{z}, M_{z}$) follows the same sequence as those
in Fig.~1(a). Figs.~2(b) and 2(c) give the corresponding
paramagnetic and diamagnetic currents, respectively. The
paramagnetic current jumps with the GS phase change while the
diamagnetic current is a linear function of the magnetic field. The
total momentum increases with increasing magnetic field with jumps
$\Delta M_z =(N\pm1)/2$. But the total spin and iso-spin remain
$S_z=1/2$ and $I_z=N/2$.

\begin{figure}[t*]
\includegraphics[angle=-90,scale=0.35]{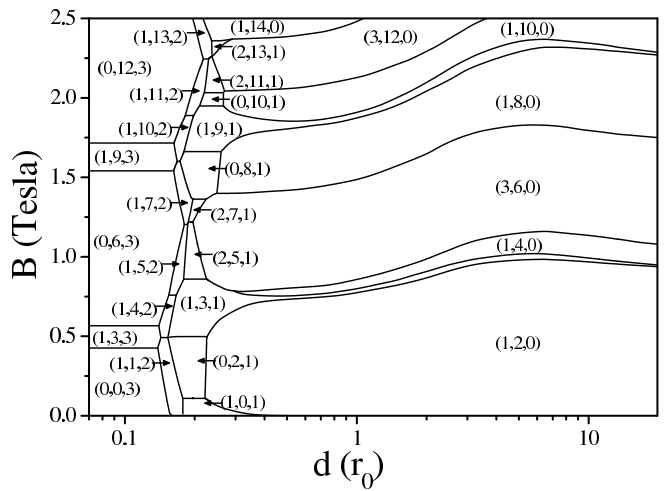}
\caption{The phase diagrams of the ground state configurations of
the CQRs of N=6 electrons. The three quantum numbers
$(S_{z},M_{z},I_{z})$ indicate the ground state
configuration.}\label{fig.3}
\end{figure}

Fig.~2(d) presents the persistent current in the CQRs of 6 electrons
at $d=0.1$~r$_0$ which is in the strong tunneling regime where
$I_z=N/2=3$. When the electron number is even ($N$=6), with
increasing magnetic field, the total angular momentum always
increase by $\Delta M_z =N/2$ when the system transits from one
phase to the next as indicated in the figure. The total spin
oscillates between 0 and 1. The most stable states correspond to the
closed-shell configurations ($S_z=0$). When the total spin is
non-zero ($S_z=1$), the ground state is less stable because it
corresponds to an open-shell system.

In the cases of a single ring and the strongly coupled double rings,
new GS phases appear with increasing magnetic field due mainly to
the increase of the total angular momentum. However, with increasing
inter-ring distance, when the tunneling splitting energy between the
bonding and antibonding levels reduces to compete with the
inter-ring exchange-correlation energy, the molecular phases can be
complex and so the persistent current.

In order to understand better the persistent current in the CQRs in
the intermediate and weak tunneling regime, we plotted in Fig.~3 the
phase diagram of the CQRs of 6 electrons (on the average 3 electrons
per ring) in the $B-d$ plane. The phase diagram of the corresponding
two ring case without tunneling was presented in Fig.~1(b). It is
seen that for small $d$, the coupled rings behave like a single one
because of the strong tunneling (large $\Delta_{\rm SAS}$). All the
electrons are in the bonding states in this case. With increasing
$d$, the coupled rings are in the so-called molecular phase. For
0.15r$_0$ $\lesssim d \lesssim 0.4$r$_0$, competition between the
tunneling energy $\Delta_{\rm SAS}$ and the electron exchange energy
leads to a quite complex phase diagram. The electrons occupy the
antibonding state but most of the electrons are in the bonding state
($I_z=1$ or 2). At large $d$, $\Delta_{\rm SAS}$ approaches zero,
the same number of electrons are in the bonding and antibonding
states, so $I_z=0$. In this case, the inter-ring direct Coulomb
interaction and the exchange energy dominate the GS phases. In
comparison with Fig.~1(b), the inter-ring tunneling alters the phase
diagram dramatically. For instance, in the phase (3,6,0), 3
electrons are in the bonding and 3 in the antibonding state. All
these 6 electrons have parallel spin. In a real sample, the
tunneling at large inter-ring separation can be controlled by a weak
external electric field in the $z$-direction.

\begin{figure}
\includegraphics[angle=0,scale=0.45]{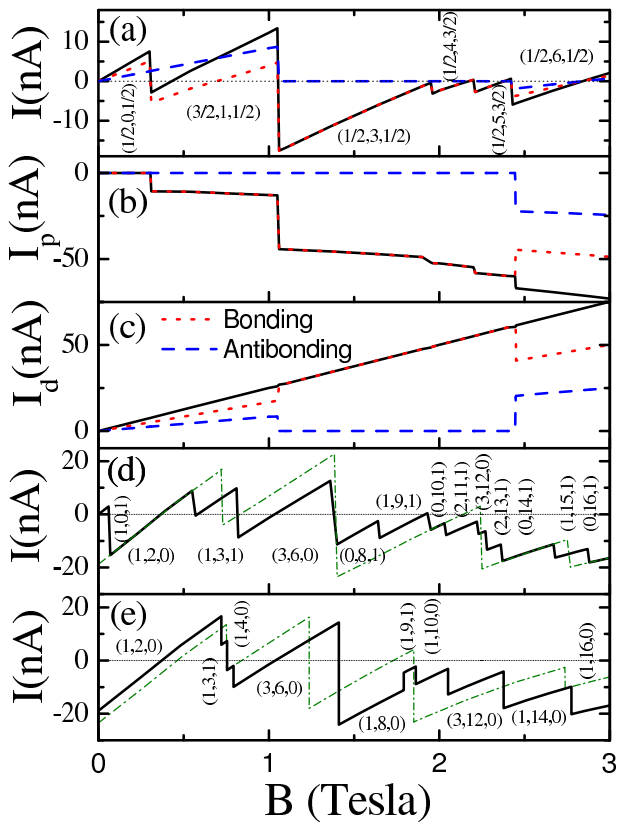}
\caption{(Color online) (a) The persistent current, (b) the
paramagnetic and (c) the diamagnetic current in the CQRs with
$d=0.23$r$_0$ and $N=$3 electrons. The dotted (red) and dashed
(blue) curves are currents carried by the bonding and antibonding
states, respectively. The persistent current in the CQRs of N=6
electrons at (d) $d=0.23$r$_0$ and (e) $d=0.40$r$_0$. The thin
dash-dotted curves are the current in the corresponding coupled
rings without tunneling. Each GS phase is indicated by
$(S_{z},M_{z},I_{z})$ }\label{fig.4}
\end{figure}

In the intermediate tunneling regime (e.g., $d=0.23$ r$_0$), the
CQRs are in the molecular phase. Fig. 4(a) shows the persistent
currents as a function of the magnetic field in the CQRs of 3
electrons. In this regime both the bonding and antibonding states
are occupied as indicated by the value of $I_z$. Therefore, more
combinations of angular momenta become possible as a function of the
magnetic field leading to more oscillations in the persistent
current. At large magnetic fields, the rapid variation of the
angular momentum with increasing magnetic field leads to a
suppression of the amplitude of the persistent current. The dotted
(red) and dashed (blue) curves indicate the currents carried by the
electrons in the bonding and antibonding states, respectively. The
black curve represents the total current, which is a sum of the
currents due to the bonding and antibonding states. Figs.~4(b) and
4(c) show the corresponding paramagnetic and diamagnetic current,
respectively. In Fig.~4(b), we see that the paramagnetic current due
to the antibonding state is zero until B=2.35T where the (1/2,6,1/2)
state appears. An increase of the current due to antibonding states
is always partially compensated by a decrease of the current due to
the bonding states. Such a compensation is complete for the
diamagnetic current as shown in Fig. 4(c).

Figs.~4(d) and 4(e) show the persistent current as a function of the
magnetic field in the CQRs of 6 electrons at $d=0.23$r$_0$ and
0.40r$_0$, respectively. The GS phase diagram of these CQRs is given
in Fig.~3. The dotted-dash (green) curves in Figs.~4(d) and 4(e)
give the current in the corresponding rings without tunneling (3
electrons in each ring). Clearly the tunneling leads to different
ground state phases and consequently different persistent current.
We notice that in the CQRs of $d=0.23$r$_0$, the persistent current
is zero at $B=0$ (see Fig.~4(d)) when there is tunneling between the
two rings. But this current is about -20 nA when the tunneling is
absent. We know that a vertical gate potential can control tunneling
in such a structure. It means that an external gate can switch
on/off the persistent current (and the magnetization) of the these
coupled rings even in the absence of a magnetic field.

\section{Conclusions}
In summary, the only external parameter which controls the
persistent current in a single QR is the external magnetic field. In
this calculation we show that in coupled quantum rings, the
persistent currents in the CQRs can be controlled by the inter-ring
coupling. This inter-ring coupling consists of the inter-ring
Coulomb interaction and the inter-ring quantum tunneling. The former
depends not only on the inter-ring distance, but also on the
electronic configuration of the QRs which can be modulated by an
external magnetic field. The later depends on the inter-ring
tunneling energy, the inter-ring electron exchange effects as well
as the external magnetic field and can also be controlled by a
vertical gate.

\acknowledgments This work was supported by FAPESP and CNPq (Brazil)
and by the Flemish Science Foundation (FWO-Vl) and the Belgium
Science Policy (IAP), Belgian State Policy.

\end{document}